\documentclass[conference]{IEEEtran}
\IEEEoverridecommandlockouts
\usepackage{cite}
\usepackage[nolist,nohyperlinks]{acronym}
\usepackage{amsmath,amssymb,amsfonts}
\usepackage{algorithmic}
\usepackage{graphicx}
\usepackage{url}
\usepackage{textcomp}
\usepackage{tikz}
\usepackage{pgfplots}
\usetikzlibrary{matrix}
\usepackage{xcolor}
\def\BibTeX{{\rm B\kern-.05em{\sc i\kern-.025em b}\kern-.08em
    T\kern-.1667em\lower.7ex\hbox{E}\kern-.125emX}}
\usepackage{pifont}
\usepackage{booktabs}
\usepackage[capitalise]{cleveref}
\usepackage{float}
\usepackage{mathtools}
\usepackage{tabularx}
\usepackage{xcolor}
\usepackage{algorithm}
\usepackage{enumitem}
\usepackage{xspace}
\usepackage{adjustbox}

\newcommand{\burn}{\textsc{burn}\xspace}
\newcommand{\claim}{\textsc{claim}\xspace}
\newcommand{\confirm}{\textsc{confirm}\xspace}

\newcommand{\tx}[1]{\ensuremath{\textit{tx}_\text{#1}}}

\newcommand{\submitter}[1]{\ensuremath{\textit{submitter}(#1)}}
\newcommand{\contract}[2]{\ensuremath{c_\textit{#1}}}

\newcounter{requirement}[subsection]
\renewcommand{\therequirement}{\arabic{requirement}}
\newcommand{\requirement}[1]{%
  \refstepcounter{requirement}
  \subsubsection*{Requirement \therequirement \:- #1}%
}

\crefname{requirement}{Requirement}{Requirements}
\Crefname{requirement}{Requirement}{Requirements}

\newfloat{protocol}{tbp}{lop}
\floatname{protocol}{Protocol}
\crefname{protocol}{Protocol}{Protocols}
\Crefname{protocol}{Protocol}{Protocols}

\newcommand{\sbline}{\\[.5\normalbaselineskip]}

\newcommand{\cmark}{\ding{51}}

\begin{document}

\title{Decentralized Cross-Blockchain Asset Transfers\\
}


\author{
 \IEEEauthorblockN{
    Marten Sigwart\IEEEauthorrefmark{1},
    Philipp Frauenthaler\IEEEauthorrefmark{1},
    Christof Spanring\IEEEauthorrefmark{2},
    Michael Sober\IEEEauthorrefmark{1},
    Stefan Schulte\IEEEauthorrefmark{1}}
 \IEEEauthorblockA{\IEEEauthorrefmark{1}\textit{Christian Doppler Laboratory for Blockchain Technologies for the Internet of Things} \\ \textit{Distributed Systems Group, TU Wien, Vienna, Austria} \\
    \{m.sigwart, p.frauenthaler, m.sober, s.schulte\}@dsg.tuwien.ac.at}
 \IEEEauthorblockA{\IEEEauthorrefmark{2}\textit{Pantos GmbH, Vienna, Austria} \\
    christof.spanring@bitpanda.com}
}

\maketitle

\begin{abstract}
Today, several solutions for cross-blockchain asset transfers exist. However, these solutions are either tailored to specific assets or neglect finality guarantees that prevent assets from getting lost in transit.

In this paper, we present a cross-blockchain asset transfer protocol that supports arbitrary assets and adheres to finality requirements. The ability to freely transfer assets between blockchains may increase transaction throughput and provide developers with more flexibility by allowing them to design digital assets that leverage the capacities and capabilities of multiple blockchains.

\end{abstract}


\begin{acronym}
	\acro{PoA}{Proof of Authority}
	\acro{PoW}{Proof of Work}
	\acro{PoS}{Proof of Stake}
	\acro{SPV}{Simplified Payment Verification}
	\acro{BAR}{Byzantine, Altruistic, Rational}
	\acro{ETH}{Ether}
	\acro{EVM}{Ethereum Virtual Machine}
	\acro{DAG}{Directed Acyclic Graph}
	\acro{DApp}{decentralized application}
	\acro{NiPoPoW}{Non-interactive Proof of Proof of Work}
	\acrodefplural{NiPoPoW}[NiPoPoWs]{Non-interactive Proofs of Proof of Work}
	\acro{ATMS}{ad-hoc threshold multisignatures}
	\acro{XCMP}{Cross-chain Message Passing}
	\acro{IBC}{Interblockchain Communication}
\end{acronym}
\section{Introduction}
\label{sec:intro}
With its ability to store data and perform computations in a decentralized and immutable manner, blockchain technology shows potential in application areas such as finance~\cite{nakamoto2008bitcoin}, supply chain management~\cite{tian2016agri}, healthcare~\cite{mettler2016blockchain}, or business process management~\cite{prybila2017runtime}. To address the diverse requirements of these areas, multiple independent and unconnected blockchains have been developed~\cite{schulte2019towards}. As it is unlikely that a single blockchain emerges that caters for the needs of all these different areas~\cite{zamyatin2019sok}, there is a strong need for interoperability between different blockchains. 

Especially in scenarios where assets, i.e., digital representations of value, are managed on-chain, the lack of interoperability leads to a vendor lock-in as assets cannot leave the blockchain platform on which they were issued. This vendor lock-in exposes projects to significant risks such as limited scalability~\cite{cai2018decentralized}, the risk that the underlying blockchain sinks into insignificance~\cite{bourgi2018fails}, and the inability to take advantage of new features offered by novel blockchains~\cite{schulte2019towards}. Of course, a centralized entity can be deployed to migrate assets from one blockchain to another, however, this contradicts the blockchain's original idea of decentralization~\cite{belchior20}. 

The ability to transfer assets to arbitrary blockchains in a decentralized way would remove the need to fully commit to a particular blockchain. Instead, assets could be migrated to new blockchains offering novel functionality or better security at any time~\cite{schulte2019towards}. Another potential use case of cross-blockchain asset transfers arises in the context of sidechains~\cite{back2014enabling, poon2017plasma}. The idea is that an asset can be transferred to and processed on multiple ``side'' blockchains, thus reducing the workload of the original blockchain.

One way to exchange assets between independent blockchains is via atomic swaps~\cite{herlihy2018atomic}. However, atomic swaps do not constitute true cross-blockchain asset transfers as no asset is transferred from one blockchain to the other but rather ownership of different assets changes in an atomic fashion without the different assets leaving their respective blockchains. True cross-blockchain asset transfers, on the contrary, achieve that an asset moves from one blockchain to the other, enabling users to hold different denominations of the same asset type on multiple blockchains.

While schemes for true cross-blockchain asset transfers have been proposed before, most of these solutions are designed with specific assets in mind and neglect important requirements for cross-blockchain asset transfers, such as finality, that prevents assets from getting lost in transit. 

Thus, in this paper, we formally define a set of general requirements that need to be fulfilled by cross-blockchain asset transfers, and then define a protocol specification that complies with the defined requirements. Additionally, we evaluate the protocol on public Ethereum test networks using a proof-of-concept implementation for EVM\footnote{Ethereum Virtual Machine}-based blockchains.

To this end, \cref{sec:background} provides important background information. In \cref{sec:approach}, we formally define the requirements and specify the protocol for cross-blockchain asset transfers. \Cref{sec:eval} evaluates the proposed protocol concerning cost, duration, security, and features using a proof-of-concept implementation. \Cref{sec:related} provides an overview of related work. Finally, \cref{sec:conclusion} concludes the paper.


\section{Background}
\label{sec:background}

This section introduces some notations and definitions necessary for describing the requirements and specification of the proposed cross-blockchain asset transfer protocol.  

As has already been mentioned in \cref{sec:intro}, cross-blockchain asset transfers ideally enable users to hold different denominations of the same asset on multiple blockchains at the same time, i.e., users are free to choose on which blockchain they want to hold their assets. An asset can be seen as anything holding some value with a corresponding representation on a blockchain. 

Assets can generally be divided into fungible and non-fungible assets~\cite{PillaiBM19}. Fungibility implies that two entities of the same asset can be used interchangeably. Cryptocurrencies like Bitcoin or Ether are fungible assets. A further example of fungible assets are Ethereum tokens following the ERC20 standard~\cite{cuffe2018role}. In contrast, non-fungible assets are uniquely identifiable, i.e., one entity cannot be simply substituted by another entity. For instance, Cryptokitties are non-fungible assets.

One can further distinguish between native and user-defined assets~\cite{PillaiBM19}. Native assets are inherently part of a particular blockchain. One cannot exist without the other, e.g., the Bitcoin and Ether cryptocurrencies and the Bitcoin and Ethereum blockchains, respectively. On the other hand, certain blockchains allow the implementation of use-case-specific assets with their own set of rules, e.g., the already mentioned ERC20 tokens~\cite{cuffe2018role}. Contrary to native assets, these user-defined assets are not bound to specific blockchains. Instead, they are implemented using smart contracts and can thus be potentially deployed on any blockchain that possesses the necessary scripting capabilities to express the asset's rules.

This work concentrates on user-defined assets since the goal is to provide an asset that allows users to hold different amounts of the same asset on multiple blockchains at the same time. We formally define an asset $A$ as a set where the set's members represent the asset's smallest indivisible entities (asset entities). For instance, the smallest indivisible entity of a fungible asset like Bitcoin is a Satoshi (i.e., 0.00000001~BTC). For a non-fungible asset like Cryptokitties, the smallest indivisible entity is a single ``cryptokitty''. Accordingly, $|A|$ represents the total supply of the asset.

We define the set of blockchains between which asset transfers can take place by the finite set $B$ (also referred to as cross-blockchain ecosystem). Each blockchain $b \in B$ can host multiple smart contracts. Out of these smart contracts, one contract is responsible for managing asset $A$ on $b$. We denote this particular smart contract as $\contract{b}{}$ for each $b \in B$.

We assume blockchains to roughly follow the model devised by Satoshi Nakatomo~\cite{nakamoto2008bitcoin}: The state of the blockchain is updated through transactions that can be used to transfer the native asset of the blockchain, to store arbitrary data, or to trigger the execution of smart contracts. In the latter case, the transaction's payload contains parameters based on which the smart contracts may change their associated state. For instance, a transaction payload containing a sender, a recipient, and an amount could trigger a smart contract causing the transfer of some user-defined asset from the sender to the recipient.

We specify transactions as a tuple containing the elements of the payload which serve as parameters for the invoked contract. In particular, we use $\tx{} \coloneqq \langle param_1, \dots, param_n \rangle$ to denote a transaction \tx{} with a payload containing $n$ parameters. Further, we define the function $\textit{calledContract}(\tx{})$ to return the address of the smart contract that was triggered by transaction~\tx{}. The execution of transactions may fail, for instance, if a user does not have enough funds for a transfer. For this, we define the function $\textit{isSuccessful}(\tx{})$ to return \textit{true} or \textit{false} depending on whether the transaction has been executed successfully or not.

Every transaction is signed by some off-chain \mbox{user $u \in U$} before being submitted to the blockchain. The function \submitter{\tx{}} denotes the user that signed \tx{}. Users can be the owners of a subset of asset $A$ on each participating blockchain~$b \in B$. Subsequently, the set $A_{u}^{b} \subseteq A$ defines the entities of asset $A$ that are owned by a particular user $u \in U$ on blockchain $b$.

Finally, for two blockchains $\textit{src}, \textit{dest} \in B$ and two users \textit{sender}, \textit{recipient} $\in U$, we define a cross-blockchain asset transfer as transfer of some $X \subseteq A$ from user \textit{sender} on source blockchain \textit{src} to user \textit{recipient} on destination blockchain \textit{dest}.

\section{Cross-Blockchain Asset Transfers}
\label{sec:approach}

In this section, we first define requirements for cross-blockchain asset transfers and then use these requirements as the foundation to define a decentralized cross-blockchain asset transfer protocol. 

\subsection{Requirements}
\label{sec:requirements}
As defined in \cref{sec:background}, a cross-blockchain asset transfer for an asset $A$ constitutes the transfer of ownership of some subset $X \subseteq A$ from some user \textit{sender} on a source blockchain \textit{src} to another user \textit{recipient} on a destination blockchain \textit{dest}. 

Before the transfer, $X$ must only exist on blockchain \textit{src}, and after the transfer, the asset must only exist on blockchain \textit{dest}. At no point should $X$ exist on both blockchains in parallel, since the accidental duplication of asset entities can potentially lead to a deflation of the asset's value. Hence, a cross-blockchain asset transfer should only be successful, i.e., $X$ is created on \textit{dest}, if $X$ has been priorly burned (i.e., destroyed) by its owner on \textit{src}.

Therefore, before $X$ can be recreated on \textit{dest}, \textit{dest} needs some kind of evidence that $X$ has already been burned on \textit{src}. If we assume that it is possible to provide such evidence guaranteeing that $X$ has been burned on \textit{src} and that this evidence can be used to recreate $X$ on \textit{dest}, two further requirements emerge. First, faking the evidence needs to be prevented at all cost. Users should not be able to counterfeit evidence certifying that $X$ has been burned on \textit{src} without it having occurred. Second, if the evidence is correct, it should only be usable once to recreate $X$ on a different blockchain, i.e., on blockchain \textit{dest}. Hence, evidence of $X$ having been burned on \textit{src} cannot be used multiple times to recreate $X$ on other blockchains. Essentially, disregard of any of these requirements would enable users to illegally create new entities of asset $A$ out of nothing---again potentially deflating the value of the asset and decreasing trust in this particular asset.

A further requirement comes up when trying to prevent the opposite, accidental inflation of the asset's value. Accidental inflation could take place if $X$ is burned on \textit{src} without ever being recreated on \textit{dest} reducing the total supply of $A$. Hence, cross-blockchain asset transfers need to be eventually finalized in order to not decrease the total supply of $A$ over time. That is, either the transfer is executed completely or it fails with no intermediate state persisting.



To sum up, we define the general requirements for a cross-blockchain asset transfer as follows:

\requirement{Ownership}
\label{req:ownership}
When a user \textit{sender} wants to burn $X$ on blockchain \textit{src}, $X$ should only be burned if $X \subseteq \nolinebreak A_{\textit{sender}}^{\textit{src}}$.

\requirement{No Claim Without Burn}
\label{req:burn-x-claim-x}
When transferring some $X \subseteq A$ from the source blockchain \textit{src} to the destination blockchain \textit{dest}, $X$ should only be recreated on \textit{dest}, if it can be proven that $X$ has already been burned on \textit{src}. That is, it should not be possible to counterfeit the burning of asset entities.

\requirement{Double Spend Prevention}
\label{req:no-double-spend}
Double spending must be prevented at all times. That is, if $X$ is burned on one blockchain, $X$ can only be recreated once on one other blockchain.

\requirement{Decentralized Finality}
\label{req:finality}
If $X$ is burned on one blockchain, $X$ is always recreated on another blockchain within a certain time limit $t$. Further,  finality should not be dependent on a single actor (i.e., not be centralized).



\subsection{Protocol}
\label{sec:protocol}


\begin{protocol*}[t]
	\caption{Protocol for cross-blockchain asset transfers}
	\label{protocol:base}

	\flushleft{
		\textit{Goal:} 
		For two blockchains $\textit{src}, \textit{dest} \in B$ and two users $\textit{sender}, \textit{recipient} \in U$, transfer $X \subseteq A$ from  \textit{src} to \textit{dest} and change ownership of $X$ from \textit{sender} to \textit{recipient}.
		\sbline
	}
	
	\begin{enumerate}[label={\arabic{enumi}.}, leftmargin=*]
		\item \textbf{Burn.} 
		User \textit{sender} creates a new \burn transaction $\tx{\burn} \coloneqq \langle \textit{recipient}, \textit{dest}, X \rangle$.
	  	\begin{enumerate}[leftmargin=*]
		\item User \textit{sender} signs and submits \tx{\burn} to source blockchain \textit{src} invoking contract \contract{src}{A}, i.e., the contract managing asset~$A$ on $src$.\label{protocol:base:txburn-submit}
	    	\item When being invoked, contract \contract{src}{A} performs the following operations.\label{protocol:base:txburn-verify}
	    	\begin{enumerate}[leftmargin=*]
	    		\item Verify $\textit{dest} \in B$ to make sure that the specified blockchain \textit{dest} is part of the cross-blockchain ecosystem.\label{protocol:base:c-verify1}
	    		\item Verify $X \subseteq A_{\textit{sender}}^{\textit{src}}$ to make sure that user \textit{sender} owns the asset entities it wants to transfer on blockchain \textit{src}.\label{protocol:base:c-verify2}
	    		\item When all checks are successful, the asset entities to be transferred are burned, i.e., $A_{\textit{sender}}^{\textit{src}} = A_{\textit{sender}}^{\textit{src}} \setminus X$.\label{protocol:base:txburn-burn}\sbline
	    	\end{enumerate}
  		\end{enumerate}
		\item \textbf{Claim.}
		Once \tx{\burn} is included in blockchain~\textit{src}, any user $\textit{u} \in U$ can construct the \claim transaction $\tx{\claim} \coloneqq \langle \tx{\burn}, \textit{proof}_{\tx{\burn}} \rangle$. Variable $\textit{proof}_{\tx{\burn}}$ contains the Merkle proof of membership of \tx{\burn} certifying the inclusion of \tx{\burn} in blockchain~\textit{src}.
	  	\begin{enumerate}[leftmargin=*]
		\item {User \textit{u} signs and submits \tx{\claim} to blockchain $b \in B$ invoking contract $\contract{b}{A}$, i.e., the contract managing asset $A$ on $b$.\label{protocol:base:txclaim-submit}}
			\item When being invoked, contract \contract{b}{A} utilizes the relay contract $\contract{relay}{A}$ to verify the inclusion and confirmation of \tx{\burn} in blockchain~\textit{src}, i.e., \contract{b}{A} calls $\contract{relay}{A}.\textit{verifyInclusion}(\tx{\burn}, \textit{proof}_{\tx{\burn}}, \textit{src})$.\label{protocol:base:txinc-verify}
			
	    	\item If $\contract{relay}{A}$ confirms the inclusion of \tx{\burn}, contract \contract{b}{A} performs the following steps.\label{protocol:base:txclaim-verify}
	    	\begin{enumerate}[leftmargin=*]
	    		\item Verify $\textit{b} = \textit{dest}$ to ensure that the executing blockchain \textit{b} is the intended destination blockchain~\textit{dest}. Note that \textit{dest}, \textit{recipient}, and $X$ are available within \contract{b}{A} as these variables are contained within the payload of \tx{\burn}. \label{protocol:base:txclaim-verify1}
	    		\item Verify $\tx{\burn} \notin T_\textit{\burn}$ where $T_\textit{\burn}$ is the set of \burn transactions that have already been used to claim entities of asset $A$ on \textit{dest}. This ensures that \burn transactions cannot be used multiple times for claiming.\label{protocol:base:txclaim-verify2}
	    		\item Verify $\textit{calledContract}(\tx{\burn}) = \contract{src}{A}$ to make sure that the contract that has been invoked by \tx{\burn} is a contract authorized for managing asset $A$ on blockchain~\emph{src}.\label{protocol:base:txclaim-verify3}
	    		\item Verify that \textit{isSuccessful}(\tx{\burn}) returns true to ensure that the execution of \contract{src}{A} has been completed without error. \label{protocol:base:txclaim-verify4}
	    		
	    		
	    		\item If $\contract{relay}{A}.\textit{confirmations}(\tx{\burn}, \textit{src}) > t$, user \textit{recipient} has not submitted \tx{\claim} within time~$t$. Hence, the user $u = \submitter{\tx{\claim}}$ that submitted \tx{\claim} receives a transfer fee $X_\textit{fee} \subseteq X$ as reward for finalizing the transfer, i.e., $A_u^{\textit{dest}} = A_u^{\textit{dest}} \cup X_\textit{fee}$. Otherwise, no fee will be paid to $u$ (i.e., $X_\textit{fee} = \emptyset$), resulting in all asset entities being transferred to \textit{recipient} (see next step).\label{protocol:base:txclaim-claim-2}
	    		
	    		\item (Re-)create the asset entities and assign ownership to user \textit{recipient}, i.e., $A_{\textit{recipient}}^{\textit{dest}} = A_{\textit{recipient}}^{\textit{dest}} \cup (X \setminus X_\textit{fee})$.\label{protocol:base:txclaim-claim-3}
	    		
	    		\item Add \tx{\burn} to the set of already used \burn transactions, i.e., $T_\textit{\burn} = T_\textit{\burn} \cup \{\tx{\burn}\}$.\label{protocol:base:txclaim-claim-4}
	    	\end{enumerate}
  		\end{enumerate}
	\end{enumerate}
\end{protocol*}

As mentioned above, a cross-blockchain asset transfer should only be successful if the asset is first burned on the source blockchain and then recreated on the destination blockchain (\cref{req:burn-x-claim-x}). This requires at least two steps, one ``burn'' step on the source blockchain, and one ``claim'' step on the destination blockchain. To verify the ``burn'' step, our protocol leverages blockchain relays which enable decentralized cross-blockchain communication. In particular, blockchain relays provide an on-chain answer of whether a certain transaction is included in the source blockchain via \ac{SPV}~\cite{nakamoto2008bitcoin}.

With this in mind, we can outline a minimal protocol for cross-blockchain asset transfers. 
The protocol consists of a \burn transaction \tx{\burn} submitted to source blockchain \textit{src} and a \claim transaction \tx{\claim} submitted to destination blockchain \textit{dest}. Further, a relay contract is used that allows the asset contract on blockchain \textit{dest} to verify the inclusion of \burn transactions in blockchain~\textit{src}. The exact specification is outlined in \cref{protocol:base}. 

Initially, some user \textit{sender} creates transaction \tx{\burn}. The payload of \tx{\burn} contains the user intended as the recipient of the transfer~(\textit{recipient}), an identifier representing the desired destination blockchain~(\textit{dest}), and the asset entities to be transferred~($X$).

User \textit{sender} then signs and submits \tx{\burn} to the source blockchain \textit{src} invoking smart contract \contract{src}{A} which manages asset $A$ on blockchain \textit{src}~(Step~\ref{protocol:base:txburn-submit}). The smart contract then verifies that the specified destination blockchain \textit{dest} is part of the cross-blockchain ecosystem (Step \ref{protocol:base:c-verify1}).
 Second, the contract makes sure that user \textit{sender} is actually the current owner of $X$ on blockchain~\textit{src}~(Step~\ref{protocol:base:c-verify2}). If both checks are successful, $X$ is burned on \textit{src} (Step~\ref{protocol:base:txburn-burn}).

Once \tx{\burn} is included in blockchain~\textit{src}, any user $u \in U$ can construct the \claim transaction \tx{\claim}. The payload of \tx{\claim} consists of transaction \tx{\burn} and a Merkle proof of membership of \tx{\burn}~($\textit{proof}_{\tx{\burn}}$) that can be used by the relay contract \contract{relay}{A} on blockchain~\textit{dest} to verify the inclusion of \tx{\burn} in blockchain \textit{src}. Note that if only the sender or only the recipient of the transfer were allowed to submit \tx{\claim}, the finality of the transfer (\cref{req:finality}) would be entirely dependent on that particular user, e.g., the user could simply decide not to submit \tx{\claim}.

User $u$ then signs and submits \tx{\claim} to some blockchain $b \in B$ invoking the contract~\contract{b}{A} managing $A$ on blockchain~$b$~(Step~\ref{protocol:base:txclaim-submit}).

By invoking the relay contract \contract{relay}{A}, the contract~\contract{b}{A} checks whether \tx{\burn} is included and confirmed in blockchain~\textit{src}~(Step~\ref{protocol:base:txinc-verify}). If the relay contract does not confirm the inclusion of \tx{\burn}, the claim request is rejected. Otherwise, contract~\contract{b}{A} performs the following steps. First, it verifies that $b$ is the intended destination blockchain \textit{dest} (Step~\ref{protocol:base:txclaim-verify1}). Second, it is verified that \tx{\burn} has not been used to claim $X$ on $b$ before~(Step~\ref{protocol:base:txclaim-verify2}). Third, if both checks are successful, contract~\contract{b}{A} verifies that the contract that burned $X$ on \textit{src} is a valid contract authorized for managing $A$ on \textit{src}~(Step~\ref{protocol:base:txclaim-verify3}). If this is the case, contract~\contract{b}{A} further checks that \tx{\burn} was successful, i.e., the execution of contract~\contract{src}{A} has been completed without any error, e.g., constraint violations~(Step~\ref{protocol:base:txclaim-verify4}). This check covers the case that in some blockchains transactions may be included even if the triggered smart contract execution was not successful. While this is the case for blockchains such as Ethereum, other blockchains may not include such transactions at all.

The above checks ensure that \tx{\claim} is only successful if the corresponding \tx{\burn} was also executed successfully. To further account for \cref{req:finality} (transfer finality), the protocol must ensure that when transaction \tx{\burn} takes place on blockchain \textit{src}, the corresponding \tx{\claim} is eventually submitted to destination blockchain~\textit{dest}. 

Usually, the incentive for transfer finalization lies with the recipient of the transfer since the recipient wants to receive the transferred asset entities. However, in case the recipient is indisposed to submit \tx{\claim} for some reason, the protocol offers an incentive in the form of a transfer fee to other users. That is, any user $u$ that successfully submits \tx{\claim} gets assigned a subset $X_{\textit{fee}} \subseteq X$ as reward~(Step~\ref{protocol:base:txclaim-claim-2}). However, to provide user \textit{recipient} with the chance to receive all entities of $X$, other users are only eligible to receive the fee if they submit \tx{\claim} after a certain time period $t$ has elapsed.

Time period $t$ is defined by the number of blocks that succeed the block containing \tx{\burn} on source blockchain \textit{src}. Hence, when being invoked by \tx{claim}, \contract{b}{A} additionally queries the relay contract~\contract{relay}{A} whether the block containing \tx{\burn} is confirmed by more than $t$ succeeding blocks. If this is the case, the time period $t$ is considered elapsed, and the user that submitted \tx{\claim} receives the transfer fee $X_\textit{fee}$, while user \textit{recipient} receives the rest $X \setminus X_\textit{fee}$. If not, user \textit{recipient} always receives the entire set $X$ (i.e., $X_{\textit{fee}} = \emptyset$), even if another user submitted \tx{\claim}~(Step~\ref{protocol:base:txclaim-claim-3}). Asset entities are \mbox{(re-)created} on blockchain~\textit{b}~($=\textit{dest}$) by incrementing the balance of the recipient (in case of fungible assets) or by copying the transferred asset's data structure from \tx{\burn} into the storage of contract~\contract{b}{A} (in case of non-fungible assets). 

Finally, to ensure that \tx{\burn} cannot be used to claim $X$ on $b$ again, \tx{\burn} is added to the set of already used \burn transactions $T_{\burn}$~(Step~\ref{protocol:base:txclaim-claim-4}).


\section{Evaluation}
\label{sec:eval}
For the evaluation, we provide a proof-of-concept implementation for EVM-based blockchains such as Ethereum and Ethereum Classic. The prototype, as well as the evaluation scripts used for obtaining the results presented in \cref{sec:quantitive-analysis} are available as an open-source project on GitHub\footnote{\url{https://github.com/pf92/x-chain-protocols}}.


\subsection{Prototype}
\label{sec:prototype}

As mentioned above, the prototype is implemented for EVM-based blockchains. The advantages of targeting EVM-based blockchains in a first proof-of-concept are twofold. First, EVM-based blockchains such as Ethereum are today among the most popular blockchains concerning \acp{DApp} and digital assets~\cite{cai2018decentralized, diangelo2019survey}. Cross-blockchain transfer capabilities for EVM-based blockchains can thus enhance the utility of a majority of available assets. The second reason is rather practical. As quite a few EVM-based blockchains exist\footnote{\url{https://crypt0.zone/dag-file-size/CLO/2926154}}, multiple blockchains can be targeted with a single implementation. However, as long as a blockchain provides sufficient scripting capabilities to implement the concepts of the protocol as well as some means of transaction inclusion verification~(e.g., via oracles or relays), the solution can be adopted beyond EVM-based blockchains.


For our analysis, we use the ERC20 token standard as asset representation. For transaction inclusion verification, the prototype leverages ETH Relay, a blockchain relay specifically targeting EVM-based blockchains~\cite{frauenthaler2020ethrelay}. 


A transaction in Ethereum consists of the fields \textit{nonce}, \textit{gasPrice}, \textit{gasLimit}, \textit{to}, \textit{value}, \textit{data}, and a signature (\textit{v}, \textit{r}, \textit{s})~\cite{wood2014ethereum}. Field \emph{data} contains the payload (e.g., the parameters for a smart contract invocation) of the transaction. Field \emph{to} contains the address of the smart contract that was invoked by the transaction~(i.e., function $\textit{calledContract}(\tx{})$ in our protocol). The submitter $\textit{submitter}(\tx{})$ of the transaction can be calculated out of the signature fields \emph{v}, \emph{r}, and~\emph{s}.

It should be noted that the transaction data does not contain information about the status of the transaction, i.e., whether the execution succeeded or failed. In Ethereum, this information is stored in another data structure, the so-called transaction receipt. For each transaction, there exists a corresponding receipt that contains among other fields any events that were emitted during the execution of the transaction and a status flag indicating the successful execution of the transaction. Thus, when evaluating the function $\textit{isSuccessful}(\tx{})$ in our protocol, the asset contract must have access to the receipt of \tx{} as well. 

In Ethereum, a block's transactions and receipts are kept in separate Merkle trees which do not contain references to each other. However, transaction and receipt are logically linked together as their position in their respective trees is identical. Both, the inclusion of the transaction and the receipt, can be verified via Merkle proofs of memberships using the respective Merkle trees.

Thus, to make sure that receipt and transaction belong together, both Merkle proofs need to be evaluated along the same search path. Hence, in our prototype, the proof data certifying the inclusion of a transaction (e.g., $\textit{proof}_{\tx{\burn}}$) not only contains a Merkle proof for the transaction itself but also for the receipt of the transaction, and the search path along which the Merkle proofs need to be evaluated.

\subsection{Requirements Analysis}
\label{sec:requirements-analysis}
This section evaluates the protocol with regard to the requirements defined in \cref{sec:requirements}. 

\requirement{Ownership}
When user \textit{sender} submits a \burn transaction \tx{\burn} invoking contract \contract{src}{A}, the contract verifies that $X \subseteq A_{\textit{sender}}^{\textit{src}}$~(see Step~\ref{protocol:base:c-verify2} of \cref{protocol:base}), thus making sure that user \textit{sender} is the owner of $X$ on \textit{src}. Hence, we consider \cref{req:ownership} as fulfilled.
	
\requirement{No Claim Without Burn}

To claim $X$ on \textit{dest}, a user \textit{u} submits a \claim transaction \tx{\claim}. As defined by the protocol, the user provides \tx{\burn} as well as some proof data in the payload of \tx{\claim}. Before recreating $X$, the asset contract \contract{dest}{A} on blockchain \textit{dest} performs several checks. 
	
First, it is verified that \tx{\burn} is included in the source blockchain~\textit{src} and confirmed by enough blocks (Step~\ref{protocol:base:txinc-verify}). Second, the protocol checks that \tx{\burn} indeed invoked asset contract \contract{src}{A} on the source blockchain~\textit{src} (Step~\ref{protocol:base:txclaim-verify3}). Third, contract~\contract{dest}{A} verifies that the execution of \tx{\burn} was successful (Step~\ref{protocol:base:txclaim-verify4}).
	
These three checks ensure that assets are created on the destination blockchain \textit{dest} if and only if they have been successfully burned by the contract \contract{src}{A} on source blockchain~\textit{src}. Notably, the fulfillment of this requirement strongly depends on the security of the used transaction inclusion verification mechanism. A security analysis of the blockchain relay \mbox{(ETH Relay)} used in our proof-of-concept implementation can be found in~\cite{frauenthaler2020ethrelay}.

\requirement{Double Spend Prevention}

To ensure that burned assets can not be claimed multiple times, all \burn transactions that have already been used within \claim transactions are stored in a set $T_\textit{\burn}$ within asset contract \contract{dest}{A}. When \contract{dest}{A} is invoked by a new \claim transaction \tx{\claim}, it only executes the claim if the provided \burn transaction \tx{\burn} is not yet included in $T_\textit{\burn}$~(Step~\ref{protocol:base:txclaim-verify2}).
	
Further, by encoding an identifier of the desired destination blockchain \textit{dest} within \burn transactions, a burned asset can not be claimed on multiple different blockchains. When an asset contract \contract{b}{A} on some blockchain~\textit{b} is invoked by a \claim transaction containing \tx{\burn}, \contract{b}{A} can verify whether it is the intended destination contract by comparing $\textit{b} = \textit{dest}$~(Step~\ref{protocol:base:txclaim-verify1}). If not, the claim is rejected. Therefore, \cref{req:no-double-spend} can be considered fulfilled as well.
	
\requirement{Decentralized Finality}

For the analysis of finality, we make use of the \acs{BAR}~(\aclu{BAR}) model~\cite{Aiyer2003BARModel} which has found application in security analysis for blockchain protocols and extensions before~(e.g.,~\cite{frauenthaler2020ethrelay,herlihy19cross, Judmayer2019PayToWinIA}). Under this model, \textit{Byzantine} users may depart arbitrarily from the protocol for any reason; \textit{altruistic} users always adhere to the protocol rules and \textit{rational} users will deviate from the protocol to maximize their profit.

The protocol in this paper offers a reward to users submitting the \claim transaction. As the reward is at least as high as the submission cost of \claim transactions~(see~\cref{sec:quantitive-analysis}), rational users have an economic incentive to finalize transfers. However, in any protocol, rational users according to the \ac{BAR}~model cannot be fully trusted to act rationally in the sense of the protocol's incentive structure since seemingly irrational behavior might be perfectly rational in the context of a larger ecosystem with the protocol being part of it~\cite{ford2019rationality}. For instance, rational users might aim at yielding profit in the larger ecosystem by finding ways to bet against the protocol or the value of the asset.

Therefore, rational users are not guaranteed to comply with the protocol rules even with perfectly aligned incentives. In fact, building an open and permissionless system that withstands all participants potentially deviating from the protocol rules appears fundamentally impossible~\cite{ford2019rationality}. Thus, in our protocol, not only the users directly involved in a transfer are allowed to post the \claim transaction, but rather any user of the system can do it. This provides stronger finalization guarantees as finalization does not depend on a single user acting honestly. Rather, it is sufficient if one user out of all users is altruistic to ensure finalization. Notably, the protocol only relies on an altruistic user in case rational users see an incentive in deviating from the proposed protocol.

\subsection{Quantitative Analysis}
\label{sec:quantitive-analysis}
This section analyzes transfer cost and duration using the developed proof-of-concept implementation. For that, we conduct cross-blockchain asset transfers between the public Ethereum test networks Rinkeby and Ropsten. 
Our evaluation consists of conducting 500 transfers of 1 ERC20 token from Rinkeby to Ropsten, i.e., \burn transactions are submitted to Rinkeby whereas \claim transactions are executed on Ropsten.


\subsubsection{Transfer Cost}

For every performed transfer, we measure the gas consumption of both transaction types. The obtained results are outlined in \cref{fig:transfer-cost}. Note that the figure contains the gas consumption for the protocol as well as the gas consumption of the blockchain relay (ETH Relay).

The total gas consumption is about 343.5 kGas (with a standard deviation of 25.81 kGas), calculated as the sum of \tx{\burn} and \tx{\claim}. With an exemplary exchange rate of about 1678.44~EUR per ETH (as of June 2021) and a gas price of 10~GWei, this results in transfer cost of about 5.77~EUR. Notably, transaction inclusion verification mechanisms may require a fee possibly increasing the overall cost. For simplicity, in our experiments, we set the fee that has to be paid to the blockchain relay for each inclusion verification to zero.

The execution of \tx{\burn} is the cheapest, followed by \tx{\claim}. The differences can be explained by the different payloads of each transaction type. As the payload of \tx{\burn} does not consist of any other transaction and proof data, less data needs to be passed to and processed by the asset contract leading to lower gas consumption. On the contrary, \tx{\claim} contains \tx{\burn} as well as proof data for \tx{\burn}. This leads to higher gas consumption. 

The gas consumption of transaction inclusion verifications depends on the concrete means of cross-blockchain communication, i.e., in our case ETH Relay. ETH Relay requires Merkle proofs of membership to be passed as proof data. As such, the gas consumption of the relay not only consists of the execution cost but also the cost of the data required for the execution. In our case, the gas consumption of the relay is higher than the gas consumption of the protocol alone. Other inclusion verification mechanisms may exhibit different gas consumptions.

\pgfplotsset{%
	,/tikz/font=\normalsize%
	,compat=1.11%
}

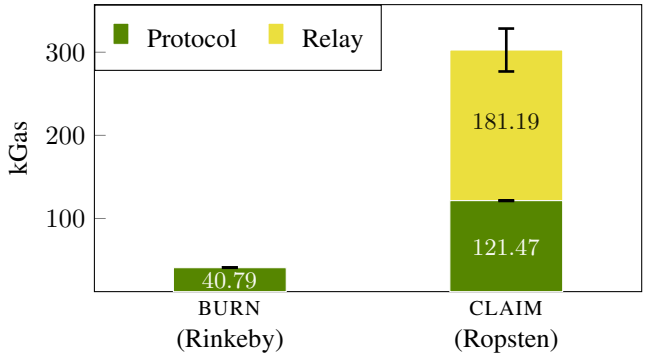
\begin{figure}[t]
	\definecolor{darkblue}{rgb}{.3375,.525,0}
	\colorlet{darkbluetext}{darkblue!5!white}
	\definecolor{darkverm}{rgb}{.96,.48,0}
	\colorlet{darkvermtext}{darkverm!30!black}
	\definecolor{dirtyelo}{rgb}{.9215,.873,.2425}
	\colorlet{dirtyelotext}{dirtyelo!10!black}
	\centering
	\begin{tikzpicture}
	\begin{axis}[
	ybar stacked,
	width = \linewidth,
	height = 5.4cm,
	bar width=1.5cm, 
	nodes near coords,
	nodes near coords style={font=\small, yshift=1.5},
	enlarge x limits={abs=1.8cm},
	ylabel={kGas},
	xtick=data,
	ytick pos=left,
	xtick pos=bottom,
	ytick={0,100,...,400},
	xticklabels = {\burn \\ (Rinkeby),\claim \\ (Ropsten),\confirm \\ (Rinkeby)},
	xticklabel style={align=center},
	]
	
	\addplot+[ybar, color=darkbluetext, draw=white, fill=darkblue, error bars/.cd, y dir=both, y explicit, error bar style={line width=1.0pt, black}, error mark options = {
		rotate = 90, 
		line width=1.0pt, 
		mark size = 3pt, 
		black}] plot coordinates {(1,40.79)+-(0,0.011) (2,121.47)+-(0,0.383)};\label{fig:transfer-cost:protocol}
	
	\addplot+[ybar, color=dirtyelotext, draw=white, fill=dirtyelo, error bars/.cd, y dir=both, y explicit, error bar style={line width=1.0pt, black}, error mark options = {
		rotate = 90, 
		line width=1.0pt, 
		mark size = 3pt, 
		black}] plot coordinates {(1,0)+-(0,0) (2,181.19)+-(0,25.81)};\label{fig:transfer-cost:relay}
	\end{axis}
	
	\matrix[
	matrix of nodes,
	anchor=north west,
	draw
	]
	at([xshift=0cm, yshift=0cm]current axis.north west)
	{
		\ref{fig:transfer-cost:protocol}& Protocol &[5pt]
		\ref{fig:transfer-cost:relay}& Relay \\
	};
	\end{tikzpicture}
	\caption{Avg. Transaction Gas Consumption}
	\label{fig:transfer-cost}
\end{figure}

\pgfplotsset{%
	,/tikz/font=\normalsize%
	,compat=1.11%
}
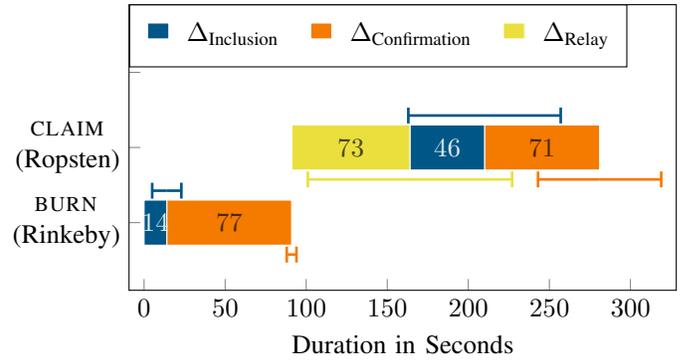
\begin{figure}[t]
	\definecolor{darkblue}{rgb}{0,.3375,.525}
	\colorlet{darkbluetext}{darkblue!10!white}
	\definecolor{darkverm}{rgb}{.96,.48,0}
	\colorlet{darkvermtext}{darkverm!30!black}
	\definecolor{dirtyelo}{rgb}{.9215,.873,.2425}
	\colorlet{dirtyelotext}{dirtyelo!30!black}
	\centering
	\begin{tikzpicture}
	\begin{axis}[
	xbar stacked,
	width = \linewidth,
	y=1.0cm, 
	bar width=0.6cm, 
	nodes near coords,
	enlarge y limits={abs=0.9cm},
	enlarge x limits={abs=0.2cm},
	xlabel={Duration in Seconds},
	ytick=data,
	ytick pos=left,
	xtick pos=bottom,
	xtick={0,50,...,400},
	yticklabels = {\burn \\ (Rinkeby),\claim \\ (Ropsten)},
	yticklabel style={align=center},
	]
	\addplot+[xbar, white] plot coordinates {(0,1) (91,2) (281,3)};

	\addplot+[xbar, color=dirtyelotext, draw=white, fill=dirtyelo, error bars/.cd, x dir=both, x explicit, error bar style={line width=1.0pt, yshift=-4.25mm, dirtyelo}, error mark options = {
		rotate = 90, 
		line width=1.0pt, 
		mark size = 3pt, 
		dirtyelo}] plot coordinates {(0,1) (73,2)+-(63,0) };\label{fig:transfer-duration:relay}
	
	\addplot+[xbar, color=darkbluetext, draw=white, fill=darkblue, error bars/.cd, x dir=both, x explicit, error bar style={line width=1.0pt, yshift=+4.25mm, darkblue}, error mark options = {
		rotate = 90, 
		line width=1.0pt, 
		mark size = 3pt, 
		darkblue}] plot coordinates {(14,1)+-(9,0) (46,2)+-(47,0)};\label{fig:transfer-duration:inclusion}
	
	\addplot+[xbar, color=darkvermtext, draw=white, fill=darkverm, error bars/.cd, x dir=both, x explicit, error bar style={line width=1.0pt, yshift=-4.25mm, darkverm}, error mark options = {
		rotate = 90, 
		line width=1.0pt, 
		mark size = 3pt, 
		darkverm}] plot coordinates {(77,1)+-(3,0) (71,2)+-(38,0)};\label{fig:transfer-duration:confirmation}
	\end{axis}
	
	\matrix[
	matrix of nodes,
	anchor=north west,
	draw
	]
	at([xshift=0cm, yshift=0cm]current axis.north west)
	{
		\ref{fig:transfer-duration:inclusion}& $\Delta_{\text{Inclusion}}$&[5pt]
		\ref{fig:transfer-duration:confirmation}& $\Delta_{\text{Confirmation}}$&[5pt]
		\ref{fig:transfer-duration:relay}& $\Delta_{\text{Relay}}$\\
	};
	\end{tikzpicture}
	\caption{Avg. Transaction Durations}

	\label{fig:transfer-duration}
\end{figure}

\subsubsection{Transfer Duration}

In this subsection, we analyze the average duration of asset transfers. After submitting a \burn transaction, the submitter may wait some time before posting the corresponding \claim transaction. Hence, the overall transfer duration depends to a large extent on the user initiating the transfer. 

Despite this uncertainty, we can measure the smallest possible transfer duration by submitting the \claim transaction at the earliest possible time, i.e., as soon as the \burn transaction is included in the blockchain and confirmed by enough blocks. In our experiment, we require each transaction on Rinkeby as well as on Ropsten to be confirmed by at least 5 succeeding blocks. Both blockchains have an inter-block time of approximately 15 seconds.

As described above, in our experiment, assets are transferred from Rinkeby to Ropsten. Therefore, \burn transactions are submitted to Rinkeby while \claim transactions are submitted to Ropsten. Hence, durations for \burn transactions were measured on Rinkeby whereas for \claim transactions on Ropsten.

Essentially, \claim transactions can be submitted to Ropsten as soon as the corresponding \burn transactions are included and confirmed on Rinkeby.  However, users need to wait until the relay running on Ropsten has been brought up to date before they can submit the corresponding \claim transactions. Otherwise, the transactions would not be successful as the relay does not have enough information to verify the inclusion of transactions yet.

To this end, $\Delta_{\text{Inclusion}}$ denotes the duration from the moment a transaction is submitted to Rinkeby (Ropsten) until it is included in some block, $\Delta_{\text{Confirmation}}$ specifies the time it takes for an already included transaction to be confirmed by enough succeeding blocks, and $\Delta_{\text{Relay}}$ denotes the time it takes for the relay to collect enough information to be able to verify the inclusion of transactions.

\Cref{fig:transfer-duration} shows the average duration for each transaction type, with the thin bars depicting the standard deviation. With an average duration of 91 seconds (standard deviation of 9~seconds), \burn transactions clearly achieve a smaller duration than \claim transactions (average duration of 191 seconds, standard deviation of 103 seconds). The total duration is calculated by summing up the durations of \burn and \claim transactions. This yields an average transfer duration of 282 seconds (standard deviation of 103 seconds).


As shown in \Cref{fig:transfer-duration}, the transfer duration depends to a large extent on the used blockchain relay, in our case ETH Relay. Other inclusion verification approaches may exhibit different durations and thus change the overall transfer duration. The transfer duration also depends on the involved blockchain's inter-block times. Hence, when being used on blockchains featuring different inter-block times than the used test networks, the average duration may change as well.

\section{Related Work}
\label{sec:related}

Several solutions for cross-blockchain asset transfers have been proposed in the literature~\cite{belchior20}. Evaluating the most important existing solutions against the requirements defined in \Cref{sec:requirements} reveals that solutions generally fulfill \crefrange{req:ownership}{req:no-double-spend}, however they lack with regards to decentralized finality~(\cref{req:finality}) or do not provide implementations of the proposed protocols. A summary of the different cross-blockchain asset transfer solutions is provided in \cref{tb:literature-review}.

In XClaim~\cite{ZamyatinHLPGK19}, cross-blockchain asset transfers are realized by first locking assets with a client called ``vault'' on a ``backing'' blockchain and reissuing the assets on another ``issuing'' blockchain. Locking of the assets on the backing blockchain is verified on the issuing blockchain via blockchain relays. However, the locked assets remain with the vault on the backing blockchain. While malicious vaults are penalized, transfer finality still depends on this single actor. In contrast, our protocol enables any client---whether directly involved in the transfer or not---to finalize transfers.

Metronome~\cite{mtn} implements cross-blockchain asset transfers for their MET token. Token holders can export MET from one blockchain and then import them on another blockchain via receipts where validators vote on the validity of receipts. While this can prevent illegal transfers, at the time of writing, Metronome cannot be considered decentralized since only authorized nodes can participate as validators. Further, Metronome does not provide concepts for transfer finality.

The authors of~\cite{KiayiasZ18} and~\cite{GaziKZ19} propose approaches for realizing cross-blockchain transfers between \ac{PoW} and \ac{PoS} blockchains, respectively. While~\cite{KiayiasZ18} verifies transaction inclusions via \acp{NiPoPoW}~\cite{kiayias2017nipopows},~\cite{GaziKZ19} enables transaction inclusion verifications via a novel cryptographic construction called \ac{ATMS}. As such, \ac{NiPoPoW} and \ac{ATMS} are used to prove events (e.g., \burn transactions) that occurred on the source blockchain to the destination blockchain. While this satisfies \cref{req:ownership,req:burn-x-claim-x}, \cref{req:no-double-spend,req:finality} are generally not covered by the protocol. Further, \acp{NiPoPoW} currently cannot be implemented in existing \ac{PoW} blockchains like Bitcoin or Ethereum without introducing a so-called velvet fork, which requires adoption from at least a subset of miners.

Similarly, Zendoo~\cite{GaroffoloKO20} provides a protocol for cross-blockchain asset transfers focussing on zero-knowledge proofs as a method for transaction inclusion verification. However, requirements such as transfer finality are not discussed. Further, the protocol relies on a special sidechain construction and can thus not be easily implemented on existing blockchains.

An approach that takes transfer finality into account is presented by van Glabbeck et al.~\cite{GlabbeekGT19}. The paper proposes a generic protocol for payments across blockchains similar to how multi-hop payment channels operate~\cite{malavolta2017concurrency}. The work has a strong focus on finality. Requirements such as double spend prevention are mentioned though not further specified. Also, the protocol has not been implemented and evaluated yet. It is difficult to tell whether the protocol allows cross-blockchain asset transfers as defined in \cref{sec:requirements} or only value transfers similar to atomic swaps.

An alternative approach for cross-blockchain asset transfers is introduced in DeXTT~\cite{borkowski2019dextt}. DeXTT describes an asset that can exist on different blockchains at the same time. However, users cannot keep different denominations of the asset on each blockchain. Rather, balances are synchronized across all participating blockchains. While the synchronization process itself is decentralized, the protocol uses a concept called claim-first transactions where assets are claimed on the other blockchains before being burned on the blockchain on which the transfer was initiated. This clearly violates~\cref{req:burn-x-claim-x}.

\newcommand{\Cm}{\cmark}
\newcommand{\Km}{\hspace{-3.5pt}(\cmark)}

\begin{table}[t]
\caption{Cross-blockchain asset transfer protocols}
\label{tb:literature-review}
\begin{flushleft}

\begin{adjustbox}{width=0.88\linewidth}
\begin{tabular}{@{}l*{6}{p{0.4cm}}}
	\textbf{Reference}         & \rotatebox{45}{Ownership} & \rotatebox{45}{No Claim Without Burn} & \rotatebox{45}{Double Spend Prevention} & \rotatebox{45}{Decentralized Finality} & \rotatebox{45}{Implementation} \\ \midrule
XCLAIM~\cite{ZamyatinHLPGK19}	& \Cm & \Cm & \Cm &     &   \Cm \\
Metronome~\cite{mtn}		& \Cm & \Cm & \Cm &     &   \Cm \\
Kiayias and Zindros~\cite{KiayiasZ18}		& \Cm & \Cm &     &     &      \\
Gazi et al.~\cite{GaziKZ19}	 	& \Cm & \Cm &     &     &      \\
Zendoo~\cite{GaroffoloKO20}	& \Cm & \Cm & \Cm &     &    \\
van Glabbeek et al.~\cite{GlabbeekGT19}		& \Cm & \Cm &     & \Cm  &     \\
DeXTT~\cite{borkowski2019dextt}	& \Cm &     & \Cm & \Cm &   \Cm \\
Proposed Protocol			& \Cm & \Cm & \Cm & \Cm &   \Cm \\
\end{tabular}
\end{adjustbox}

\end{flushleft}
\end{table}

Other works~\cite{herlihy2018atomic, herlihy19cross, interledgerprotocol15} focus on the transfer of value across different blockchains. However, these solutions rather focus on atomic swaps where two different assets are exchanged and do not constitute true cross-blockchain asset transfers as defined by our requirements in \cref{sec:requirements}.

Finally, projects such as Polkadot~\cite{polkadot} and Cosmos~\cite{cosmos} also aim for generic cross-blockchain interactions. Polkadot implements a \ac{XCMP} protocol that enables two separate parachains to communicate with each other. To accomplish this, it makes use of a simple queuing mechanism based on Merkle trees. Cosmos on the other hand implements the \ac{IBC} protocol proposed in ~\cite{goes2020interblockchain}. The \ac{IBC} protocol is inspired by the TCP/IP protocol and enables the communication between separate ledgers which implement the same interface. Multiple ledgers can establish a connection with each other to create channels over which packages can be transmitted to modules (e.g. smart contracts) on the other ledger. Both protocols have already been implemented by the respective projects.

While cross-blockchain asset transfers are mentioned as example use cases, the documentation does not mention specifics on how these transfers are implemented. Further, these projects aim to provide interoperability primarily between specialized blockchains that adhere to certain structures and consensus protocols. While there are plans to integrate existing blockchains such as Bitcoin or Ethereum into the systems, the documentation does not yet provide specifics.

To the best of our knowledge, this work is the first to provide requirements, a specification, and a proof-of-concept implementation of a cross-blockchain asset transfer protocol that also takes transfer finality into account.

\section{Conclusion}
\label{sec:conclusion}
Decentralized cross-blockchain asset transfers are one way to provide interoperability between blockchains. In particular, they prevent vendor lock-in by allowing blockchain assets to be moved away from the blockchains on which they were originally issued in a completely decentralized way. While many different solutions for enabling cross-blockchain asset transfers have been proposed, these solutions often focus on specific assets and neglect the fundamental functionality that cross-blockchain asset transfers should offer. In this work, we defined general requirements and specifications for cross-blockchain asset transfer protocols. Providing a proof-of-concept implementation of the proposed protocol, we have shown that requirements such as decentralized finality can be fulfilled.

In future work, we will investigate how the concepts of this work can be extended to provide interoperability beyond cross-blockchain asset transfers, e.g., generic message passing between blockchains.

\section*{Acknowledgment}

The financial support by the Austrian Federal Ministry for Digital and Economic Affairs, the National Foundation for Research, Technology and Development as well as the Christian Doppler Research Association is gratefully acknowledged.

\bibliographystyle{IEEEtran}
\bibliography{refs.bib}

\end{document}